\documentclass[letterpaper, 10 pt, journal, twoside]{IEEEtran}
\usepackage{microtype}

\usepackage{amsmath}
\usepackage{amsfonts}
\usepackage{amssymb}

\usepackage[]{graphicx}
\usepackage{caption}
\usepackage{multicol}
\usepackage{diagbox}
\usepackage{booktabs}
\usepackage[table]{xcolor}

\usepackage[normalem]{ulem}
\newcommand\migst{\bgroup\markoverwith{\textcolor{orange}{\rule[+0.5ex]{2pt}{3pt}}}\ULon}

\def\Real{\mathbb{R}}

\def\ie{i.e.}

\def\Fig{Fig.\,}
\def\Eq{Eq.\,}
\def\Sec{Sec.\,}

\usepackage{tikz}
\usetikzlibrary{arrows,calc}
\usepackage[]{pgfplots}

\tikzset{
    >=stealth',
    punkt/.style={
           rectangle,
           rounded corners,
           draw=black, very thick,
           text width=6.5em,
           minimum height=2em,
           text centered},
    pil/.style={
           ->,
           thick,
           shorten <=2pt,
           shorten >=2pt,}
}

\def\url{}

\usepackage{textpos}

\begin{document}

\title{Formal Synthesis of Lyapunov
  Neural Networks}

\author{Alessandro Abate, 
Daniele Ahmed, 
Mirco Giacobbe, and 
Andrea Peruffo
\thanks{This manuscript appeared on IEEE Control Systems Letters \protect \cite{journal}.
  \protect This work was supported is in part by the HICLASS project
(113213), a partnership between the Aerospace Technology
Institute (ATI), Department for Business, Energy \& Industrial
Strategy (BEIS) and Innovate UK.}
\thanks{A. Abate, M. Giacobbe, and A. Peruffo are with the Department of Computer Science,
        University of Oxford, Oxford, UK 
        (email: name.surname@cs.ox.ac.uk). }
      \thanks{D. Ahmed is with Amazon Inc., London, UK.}
}

\maketitle
\thispagestyle{empty}
\pagestyle{empty}

\begin{abstract}
  We propose an automatic and formally sound method for synthesising Lyapunov
functions for the asymptotic stability of autonomous non-linear systems.
Traditional methods are either analytical and require manual effort or are
numerical but lack of formal soundness. Symbolic computational methods
for Lyapunov functions, which are in between, give formal guarantees
but are typically semi-automatic because they rely on the user to
provide appropriate function templates. We propose a method that finds
Lyapunov functions fully
automatically---using machine learning---while also providing formal
guarantees---using satisfiability modulo theories (SMT). We employ a
counterexample-guided approach where a numerical learner and a symbolic 
verifier interact to construct provably correct Lyapunov neural networks
(LNNs). The learner trains a neural network that satisfies the Lyapunov
criteria for asymptotic stability over a samples set; the verifier proves
via SMT solving
that the criteria are satisfied over the whole domain or augments the
samples set with counterexamples. Our method supports neural
networks with polynomial activation functions and multiple depth and width,
which display wide learning capabilities. We demonstrate our method over
several non-trivial benchmarks and compare it favourably against a numerical
optimisation-based approach, a symbolic template-based approach, and a
cognate LNN-based approach. Our method synthesises Lyapunov functions faster
and over wider spatial domains than the alternatives, yet providing stronger or
equal guarantees.

\end{abstract}


\section{Introduction}
\label{sec:intro}

\IEEEPARstart{S}{tability}
 analysis determines whether a 
dynamical system never escapes a domain of interest
around an equilibrium point and, possibly, converges
asymptotically towards the point.
Stability properties constitute a primary objective
for control engineering,  
yet designing controllers for systems
that are highly complex is error prone.
Automatic stability analysis computes
certificates of stability whose aim is providing correctness guarantees to the
traditional workflow. 
We address the stability analysis of systems with given
controllers or, more generally, autonomous systems
described by non-linear ordinary differential
equations (ODEs). In particular, we present a novel method
for the automated and formal synthesis of Lyapunov
functions. 

Lyapunov functions are formal certificates for the {\em asymptotic stability}
of ODEs.
We consider autonomous $n$-dimensional systems of non-linear ODEs
\begin{equation}
  \dot{x} = f(x), \quad x \in \mathbb{R}^n, \label{eq:sys}
\end{equation}
having an equilibrium point at $x_e$ 
and a domain of interest $\mathcal{D} \subseteq \mathbb{R}^n$ containing $x_e$.  
A Lyapunov function is a real-valued function $V : \mathbb{R}^n \to \mathbb{R}$ 
such that $V(x_e) = 0$ and, for all states $x \in \mathcal{D}$ other than $x_e$,
it satisfies the two conditions
\begin{align}
  \dot{V}(x) = \nabla V(x) \cdot f(x) < 0,
  && V(x) > 0.\label{eq:lyap}
\end{align}
A Lyapunov function maps system states $x$ into
energy-like values that,
by the first condition, decrease over time along the
model's trajectories and, 
by the second condition, are bounded from below. 
If one such function exists, then the system is asymptotically
stable within $\mathcal{D}$.

Finding a Lyapunov function is in general a hard problem and has been the objective of numerous studies. 
In standard literature Lyapunov functions are constructed via analytical methods,
which are mathematically sound but require substantial expertise and
manual effort. 
Algorithmically, for linear ODEs it is sufficient
to use quadratic programming, 
as Lyapunov functions are necessarily quadratic polynomials.
However, for non-linear ODEs no general method to automatically construct Lyapunov functions exists \cite{giesl2015review}.

Numerical methods for non-linear autonomous systems include techniques that
reduce the problem to solving
a partial differential equations (PDEs), 
partition and linearise the vector field $f$ and then reformulate the problem as a linear program (LP), 
or restrict $V$ to be a sum-of-squares (SOS) function and relax the synthesis problem into a 
linear matrix inequalities (LMI) program \cite{papachristodoulou2018sostools}. 
Despite their analytical exactness,  
PDE-based methods rely on numerical integrators which are bound to machine precision,
LP-based methods linearise $f$ with finite accuracy,
and LMI-based methods employ numerical convex optimisation---unfortunately, all these methods are numerically unsound. 
Conversely, we deal with constructing a Lyapunov function as a problem of formal synthesis,
which is not only {\em automatic}, 
but also formally {\em sound}. 

Formal methods for the synthesis of Lyapunov functions
guarantee the formal correctness of their result
using satisfiability modulo theories (SMT) or a 
computer algebra system (CAS).
Typically, formal methods assume $V$ to be given in some
parameterised form, \ie, a {\em template} (a.k.a. sketch), 
and either relax the entire problem into a computationally tractable abstraction
or incrementally construct and check candidates
in a
counterexample-guided inductive synthesis (CEGIS) fashion \cite{solar2006combinatorial}.
Relaxation-based methods typically assume polynomial templates
and reformulate the problem as a semi-algebraic one \cite{she2009semi,she2013discovering} 
or as a linear program \cite{RatschanS10,sankaranarayanan2013lyapunov,ben2015linear} and solve them using a CAS or SMT; 
notably, Darboux-based semi-algebraic methods can also
relax problems with transcendental functions \cite{goubault2014finding}.
Alternatively, incremental methods construct, from polynomial templates,
candidates for $V$ using linear relaxations \cite{RS15}, 
genetic algorithms \cite{verdier2018formal},
fitting simulations or, more directly, spatial samples \cite{kapinski2014simulation,tacas20}; 
then, they verify the candidates using a CAS or SMT and,
whenever necessary, refine the search space by learning
from generated counterexamples.
Notably, all methods rely on the user to provide a good template
expression.
We overcome the limit of manually selecting a template 
using, instead of fixed expressions, generic templates
based on neural networks. 

Neural networks are widely used in a variety of applications, 
such as in image classification and in natural language processing. 
Neural networks are powerful regressors and thus
lend themselves to the approximation of 
Lyapunov functions \cite{long1993feedback,prokhorov1994lyapunov}. 
The construction of {\em Lyapunov neural networks} (LNNs) has been previously studied by approaches based on simulations and numerical optimisation, all of which are formally unsound \cite{serpen2005empirical,petridis2006construction,noroozi2008generation,richards2018lyapunov,MittalM20}.

We introduce a method that exploits 
efficient machine learning algorithms, while
guaranteeing formal soundness.
We follow a CEGIS procedure,
where first a numerical {\em learner} trains an LNN candidate
to satisfy the Lyapunov
conditions (\Eq \eqref{eq:lyap}) over a samples set and
then a formal {\em verifier} confirms or
falsifies whether the conditions are satisfied over the whole dense domain.
If the verifier falsifies the candidate, one or more
counterexamples are added to the samples set and the network is retrained.
The procedure repeats in a loop until the verifier confirms the LNN.
Our learner trains neural networks with multiple layers and
polynomial activations functions of any degree;
on the technical side, learning enjoys better performance
when the last layer has quadratic activation.
Our verifier guarantees the formal correctness of the results
using a sound decision procedure for SMT over theories for
polynomial constraints \cite{de2008z3,JovanovicM12}.
Besides, the previous CEGIS methods for LNNs provide
weaker guarantees,
namely Lagrange (practical) stability, which excludes a
neighbourhood around the equilibrium $x_e$
\cite{ChangRG19}---conversely, our novel method guarantees full
asymptotic stability at $x_e$.  

We have built a prototype software and compared our method
against a numerical LMI-based method (SOSTOOLS) \cite{papachristodoulou2018sostools},
a formal template-based CEGIS method \cite{tacas20},
and the cognate CEGIS approach for LNNs \cite{ChangRG19}. 
We have evaluated their performance over four systems of polynomial
ODEs that are challenging as do not admit polynomial
Lyapunov functions over the entire $\Real^n$.
We have thus measured the widest domain for which each of the methods succeeded to find a Lyapunov function. 
Our method has attained comparable or
wider domains than the other approaches,
in shorter or comparable time.
Notably, our method gives the strongest guarantees within the alternatives 
(asymptotic stability) and does not rely on user hints. 

Altogether, we present a synthesis method for LNNs that
(i) accounts for the asymptotic stability of systems of
non-linear ODEs,
(ii) is sound and automatic, and (iii) is faster
and covers wider domains than other state-of-the-art tools. 

\section{Counterexample-guided Inductive Synthesis of Lyapunov Neural Networks}
\label{sec:CEGIS-LNN}

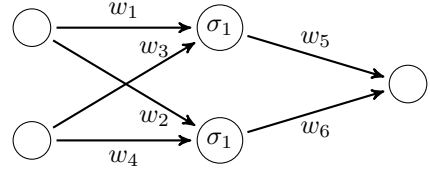
\begin{figure}
  \centering
  \begin{tikzpicture}

    \node (x1) [ draw, circle, inner sep=2pt, minimum size=0.5cm]
    at (0,1.5) {};
    \node (x2) [ draw, circle, inner sep=2pt, minimum size=0.5cm]
    at (0,0) {};
    
    \node (z1) [ draw, circle, inner sep=2pt, minimum size=0.5cm]
    at (2.5,1.5) {$\sigma_1$};
    \draw (2.75,1.85) [thick,red];
    \node (z2) [ draw, circle, inner sep=2pt, minimum size=0.5cm]
    at (2.5,0) {$\sigma_1$};
    \draw (2.75,-0.15) [thick,red];

    \node (y) [ draw, circle, minimum size=0.5cm]
    at (5,.75) {};

    
    \draw [pil] (x1) edge node[above] {$w_1$} (z1);
    \draw [pil] (x1) edge node[below, pos=0.7] {$w_2$} (z2);
    \draw [pil] (x2) edge node[above, pos=0.7] {$w_3$} (z1);
    \draw [pil] (x2) edge node[below] {$w_4$} (z2);

    \draw [pil] (z1) edge node[above] {$w_5$} (y);
    \draw [pil] (z2) edge node[below] {$w_6$} (y);

  \end{tikzpicture}
  \caption{A feed-forward neural network with one hidden layer.}
  \label{fig:lnn}
\end{figure}
We introduce a CEGIS procedure for the construction of
Lyapunov functions in the form of feed-forward neural networks.
We consider a network with a number $n$ of input neurons that corresponds
with the dimension of the dynamical system,
followed by $k$ hidden layers with respectively $h_1, \dots, h_k$ neurons,
and finally followed by one output neuron.
Nodes of adjacent layers are fully interconnected: 
a matrix $W_1 \in \mathbb{R}^{h_1 \times n}$ encompasses the weights from input to first hidden layer,
a matrix $W_i \in \mathbb{R}^{h_i \times h_{i-1}}$ the weights from any other $(i-1)$-th to $i$-th hidden layer,
and a matrix $W_{k+1} \in \mathbb{R}^{1 \times h_k}$
the weights from $k$-th layer to the last neuron.
Neurons have no additive bias.
Every $i$-th hidden layer comes with a non-linear activation function $\sigma_i \colon \mathbb{R} \to \mathbb{R}$
and the output neuron is activation free.
The valuation of output and hidden layers
are given by 
\begin{align}
  z_{k+1} = W_{k+1} z_k, && z_i = \sigma_i(W_i z_{i-1}), && i = 1, \dots, k,
\label{eq:layer}
\end{align}
where each $\sigma_i$ is applied element-wise to its
$h_i$-dimensional argument and $z_0$ is the input layer.
Upon assigning the argument $x \in \Real^n$ to the input layer,
the neural network evaluates \Eq \eqref{eq:layer} layer by layer,
resulting in the function  
\begin{align}
  V(x) = z_{k+1}, && z_0 = x.
\end{align}
Figure \ref{fig:lnn} depicts a neural network of this kind 
with $k=1$, $n=h_1=2$, and the weights $w_1, \dots, w_6$.
Unlike the standard definition, 
we assume here to have no additive bias and 
require $\sigma_i(0) = 0$ for $i = 1, \dots, k$, which results in the condition $V(0) = 0$. 

\begin{figure}[b]
  \centering
  \begin{tikzpicture}
    \node[shape=rectangle,draw,minimum width=1.7cm, minimum height=1cm] (lea) at (0,0) {Learner};
    \node[shape=rectangle,draw,minimum width=1.7cm, minimum height=1cm] (ver) at (3.5,0) {Verifier};
    \draw[->] (lea) to node[above] {$V$} (ver);
    \draw[->] (ver) to node[right] {\tt cex $C$} (3.5,-1) to node[above] {$S \leftarrow S \cup C$} (0,-1) to (lea);

    \node[align=right,text width=1.5cm] (inp) at (-2,0) {$h_1, \dots, h_k$};
    \draw[->] (inp) to ($(lea.west)+(0,0)$);

    \node[] (prob) at (1.75,1.5) {$f, \mathcal{D}$};
    \draw[->] (prob) to (1.75,1) to (0,1) to (lea);
    \draw[->] (1.75,1) to (3.5,1) to (ver);
    
    \node (out) at (5.5,0) {$V$};
    \draw[->] (ver) to node[above,xshift=.2cm,yshift=.1cm] {\tt valid} (out);
  \end{tikzpicture}
  \caption{CEGIS architecture for the synthesis of LNNs.}
  \label{fig:cegis}
\end{figure}
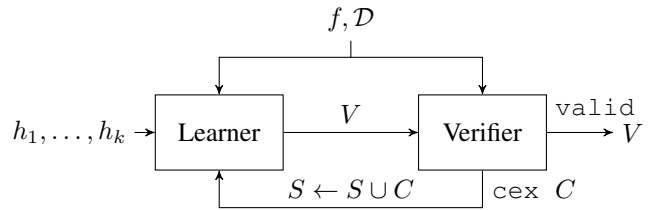 
Our procedure takes as input an
$n$-dimensional vector field
$f \colon \Real^n \to \Real^n$ with equilibrium point (w.l.o.g.)
$x_e = 0$, a domain $\mathcal{D} \subseteq \Real^n$, and
the desired depth $k$ and width $h_1, \dots, h_k$ for
the hidden layers of the network.
Upon termination, the procedure returns a 
neural network $V \colon \mathbb{R}^n \to \mathbb{R}$
that satisfies the Lyapunov conditions in \Eq \eqref{eq:lyap},
which is an LNN for the asymptotic stability of $f$ within region $\mathcal{D}$.

\begin{figure*}[t]
  \def\guessscale{.33}
  \centering
  \begin{tabular}{ccc}
    \includegraphics[scale=\guessscale]{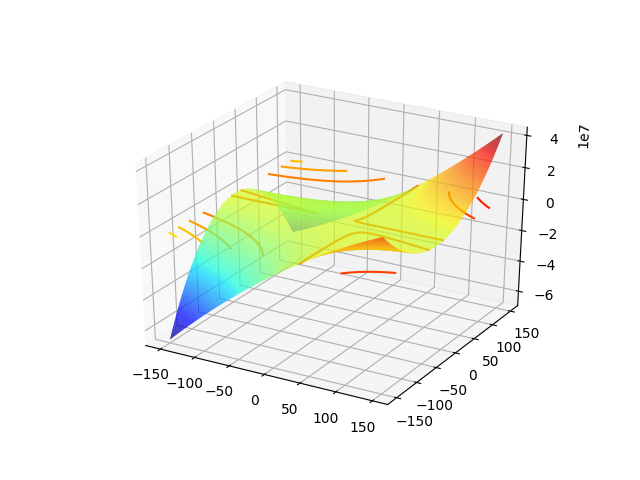}
    &
    \includegraphics[scale=\guessscale]{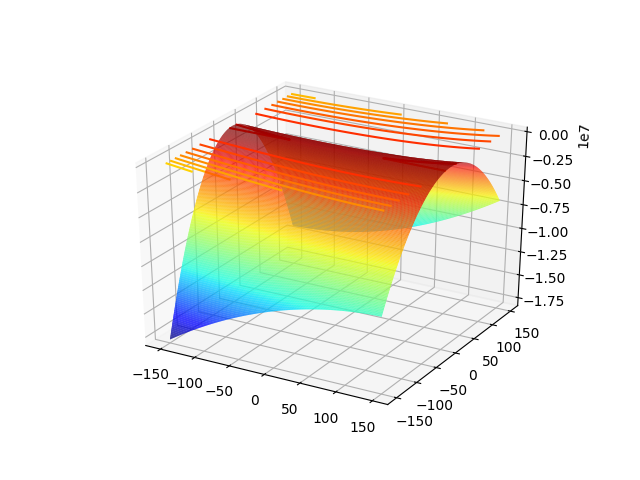}
    &
    \includegraphics[scale=\guessscale]{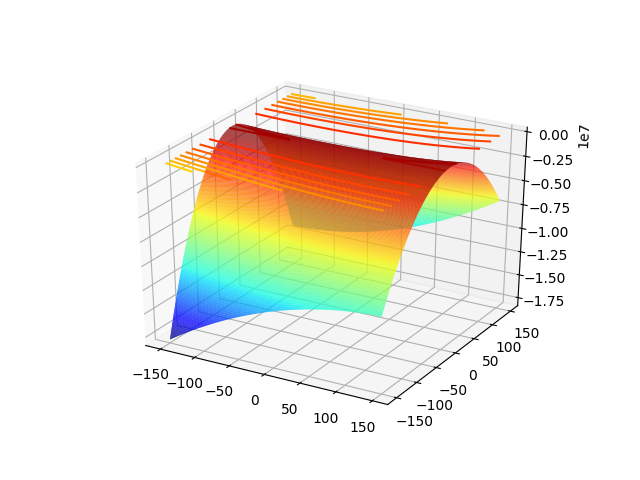}
    \\
    \includegraphics[scale=\guessscale]{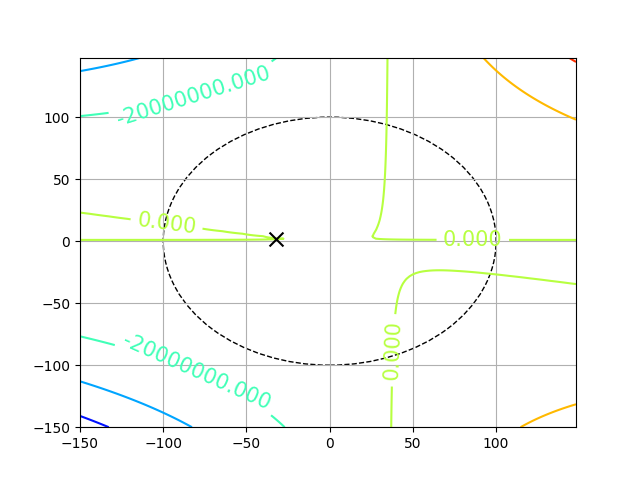}
    &
    \includegraphics[scale=\guessscale]{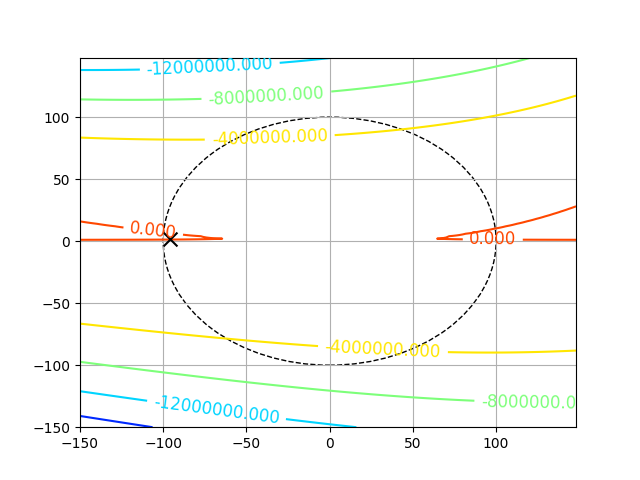}
    &
    \includegraphics[scale=\guessscale]{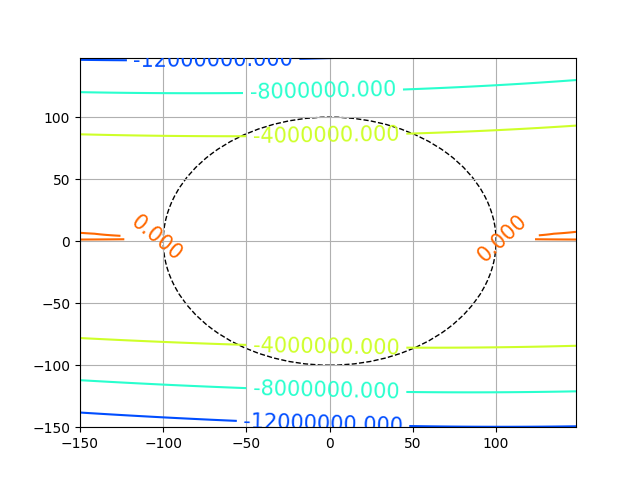}
    \\
    Initial guess. & After the first counterexample. & After the second counterexample.
    \\
    (a) & (b) & (c)
  \end{tabular}
  \caption{The evolution of $\dot{V}(x, y)$ with the corresponding level sets
    through three CEGIS iterations for certifying the asymptotic stability
    of the system in \Eq \ref{eq:parrilo} within a circle of radius $\gamma = 100$,
    using the neural network in \Fig \ref{fig:lnn}.
    The synthesis loop finds two counterexamples, shown as crosses,
    and succeeds after three iterations.
  }
\label{fig:lyap-guess}
\end{figure*}

Figure \ref{fig:cegis} outlines the architecture which
consists of a learner and a formal verifier interacting
in a CEGIS loop. 
The learner trains a candidate neural network $V$ to satisfy 
the conditions in \Eq \eqref{eq:lyap} over a discrete set of samples $S \subset \mathcal{D}$,
which is initialised randomly. 
The outcome from the learner satisfies $V(0) = 0$, $\dot{V}(s) < 0$, and $V(s) > 0$ over all samples $s \in S$,
but not necessarily over the entire dense domain $\mathcal{D}$. 
Thus the formal verifier checks whether the resulting $V$ violates the same conditions within the whole $\mathcal{D}$ and, if so, produces a set of samples $C \subset \mathcal{D}$ containing one or
more counterexamples $c$ that violate either $\dot{V}(c) < 0$ or $V(c) > 0$.
We add $C$ to the samples set $S$, 
hence forcing the learner to newly produce a different candidate function, 
which will later be passed again to the verifier.  
The loop repeats indefinitely, until the verifier proves that no
counterexamples exist: this outcome proves
that $V$ is an LNN over the entire $\mathcal D$.  
We cannot however guarantee termination of this procedure in general, 
rather we are interested in its performance in practice. 

\paragraph*{Example}
We demonstrate the workflow of our procedure
with the planar dynamical system described by 
\begin{equation}
\left\{
  \begin{array}{l}
    \dot{x} =  -x+xy\\
    \dot{y} =  -y.
  \end{array}\right.
  \label{eq:parrilo}
\end{equation}
The system is asymptotically stable at the origin
\cite{ahmadi2011globally}.
We aim at proving its stability within the circle of
radius 100 centred at the origin, that is
$\mathcal{D} = \{x \colon ||x||_2 \leq \gamma \}$
for $\gamma = 100$.
First, we select the neural network in \Fig \ref{fig:lnn},
for which $k=1$ and $h_k = 2$, and use the quadratic
activation function $\sigma_1(x) = x^2$. 
Second, we also impose $w_5 = w_6 = 1$, 
which makes the training of $V$ 
faster (see \Sec \ref{sec:learner}).
Finally, our CEGIS procedure trains a provably correct LNN
after three learner/verifier iterations; 
Figure \ref{fig:lyap-guess} shows the evolution of
$\dot{V}$ after each iteration.
At the beginning, the procedure samples a set $S$ of
random points from $\mathcal{D}$ and then invokes the learner.
The learner keeps the samples in state space fixed,
while it searches over the parameter space
$w_1, \dots, w_4$ using numerical gradient descent.
In particular, it computes a network candidate that satisfies
the Lyapunov conditions for all random initial points: 
the result is shown in \Fig \ref{fig:lyap-guess}a. 
Next, the verifier fixes the current instance of parameters
$w_1, \dots, w_4$ as constants, and the SMT solver accepts a first-order
logic formula whose variables are the state-space points   
$x \in \mathcal{D}$ that violate the Lyapunov conditions. 
The solution returned by the SMT solver is the counterexample $c$ that is
depicted in \Fig \ref{fig:lyap-guess}a as a cross, for which
$\dot{V}(c) \geq 0$. 
At the second iteration, the counterexample is added to $S$
and the network retrained over the extended batch,
obtaining the $\dot{V}$ of \Fig \ref{fig:lyap-guess}b.
Now the network satisfies the conditions over 
all initial samples plus the newly added point,
yet it violates it over a different counterexample,
which is depicted in \Fig \ref{fig:lyap-guess}b.
The verifier identifies this counterexample and adds it to $S$.
At the third and last iteration, the learner retrains
the neural network, which yields \Fig \ref{fig:lyap-guess}c.
The verifier re-checks it, but this time it fails
at producing any counterexamples, thus proving their absence. 
Consequently, the neural network satisfies the 
Lyapunov conditions over the entire continuous domain $\mathcal{D}$,  
and the CEGIS loop terminates successfully. \hfill $\square$ 

The formal synthesis of LNNs consists of
finding an instance of weights for which
the neural network satisfies the Lyapunov conditions 
of \Eq \eqref{eq:lyap}.
Our CEGIS loop tackles this general problem by solving
two separate problems interactively: 
the first is learning and the second is verifying. 
We capitalise on the power of neural networks 
for learning from data (see \Sec \ref{sec:learner}), and on
the power of SMT solving for verifying or for producing  
counterexamples accordingly (see \Sec \ref{sec:verifier}). 

\section{Training of Lyapunov Neural Networks}
\label{sec:learner}

The first active CEGIS component is the learner, 
which uses gradient descent to train LNN candidates. 
The learner instantiates a candidate using the
  hyper-parameters $k$ and $h_1, \dots, h_k$ (depth and width of the network), trains it over the discrete set of
  samples $S$, and refines its training
  whenever the verifier adds counterexamples.

The training procedure performs the minimisation of a loss function that depends on $V$ and $\dot{V}$, 
both evaluated on the data points in $S$. 
The Lyapunov requirements split the sample set $S$ into two partitions $S^-$ and $S^+$, 
such that all points $s \in S^-$ satisfy both conditions $\dot{V}(s) < 0$ and $V(s) > 0$, 
whereas all data points $s \in S^+$ violate either of them. 
The loss function should penalise all data points in the $S^+$partition, 
while rewarding the $S^-$ partition.  
To this end, we employ the Leaky ReLU function, 
which is defined as 
\begin{equation}
LR(p, a) = 
\begin{cases}
p   	& \text{if } p \geq 0\\
a p 	& \text{otherwise,}
\end{cases}  
\end{equation}
where $a$ is a (small) positive constant and $p$ is the variable of interest. 
We thus minimise the sum of LR over the values $p_1 = \dot{V}(s)$ and $p_2 = -V(s)$, as in \eqref{eq:loss-function}. 
Additionally, to enhance the numerical stability of the training, we apply a small offset $\varepsilon$ to $p_1$ and $p_2$, 
therefore rewarding data points $s$ where
$\dot{V}(s) \leq - \varepsilon$ and $V(s) \geq \varepsilon$,
and penalising them otherwise. 
%
Altogether, our loss function is 
\begin{equation}
\mathcal{L}(s) =
    \underbrace{LR(\dot{V}(s)+\varepsilon, a)}_{\mathcal{L}_1}
    + 
    \underbrace{ LR(-{V}(s)+\varepsilon, a)
}_{\mathcal{L}_2}
\label{eq:loss-function}
\end{equation}
where $\mathcal{L}_1$ accounts for training $\dot{V}(x) < 0$,
  $\mathcal{L}_2$ accounts for training $V(x) > 0$,  
and where $a$ and $\varepsilon$ are hyper-parameters defined above. 
In contrast to a standard ReLU formulation,
a Leaky ReLU rewards $S^-$ by $a$, 
which induces training below zero, 
improves learning, and
yields a numerically robust LNN candidate. 

We evaluate the expression of $\dot{V}(x) = \nabla V(x) \cdot f(x)$ directly from the matrices $W_i$, 
thus avoiding a symbolic differentiation of $V(x)$. 
Let us recall the value of the $i$-th layer,   
$z_i = \sigma_{i}(W_i z_{i-1})$, so that $z_0 = x$, whereas the output layer is activation free, hence $z_{k+1} = W_{k+1} z_{k}$. 
To compute the gradient of $V(x)$ over $x$ we use the chain rule 
\begin{equation}
\nabla V(x) = \frac{\partial V}{\partial x} = 
\prod_{i=1}^{k+1} \frac{\partial z_{i}}{\partial z_{i-1}}.
\end{equation}
After a few algebraic steps, the factors result in
\begin{equation}
\frac{\partial z_{i}}{\partial z_{i-1}} = 
\frac{\partial\sigma_{i}(W_{i} z_{i-1})}{\partial z_{i-1}}  = 
\text{diag}[\sigma_{i}' (W_{i} z_{i-1})] \cdot W_{i},
\end{equation}
for $i = 1,\dots,k$, whereas for the last layer $\partial z_{k+1}/\partial z_k = W_{k+1}$; $\sigma_i'$ is the full derivative of function $\sigma_i$ and $\text{diag}[v]$ represents a diagonal matrix whose entries are
the elements of vector $v$.
Finally, 
the gradient results in
\begin{equation}
\nabla V(x) = W_{k+1} \cdot \prod_{i=1}^{k} \text{diag}[ \sigma'_{i} (W_{i} z_{i-1}) ] \cdot W_{i}.  
\end{equation}
We compute the values of $z_i$ recursively from $z_0 = x$ using \Eq \eqref{eq:layer}.
For every point $s$ in ${S}$, we thus evaluate $\nabla V(s)$
using simple matrix-vector operations, and, along
with the value of $f(s)$,
 finally obtain $\dot{V}(s) = \nabla V(s) \cdot f(s)$.

  Training benefits from candidate networks that
  satisfy or likely satisfy one of the Lyapunov
  conditions $\dot{V}(x) < 0$ or $V(x) > 0$ a priori. 
  An example are neural networks for which the last hidden layer has
  quadratic activation and positive output, 
  \ie, $\sigma_k(x) = x^2$ and $W_{k+1} > 0$.
  For a generic selection of weights,
  these networks are likely to satisfy $V(s) > 0$
  over the samples $s \in S$.
  As a result, 
  the component $\mathcal{L}_2$
  becomes negligible
  with respect
  to $\mathcal{L}_1$ during most of the training.
  Imposing these simple conditions to the network
  improves the overall training
  performance considerably.

\begin{table*}[ht]
\centering
\begin{tabular}{c|c|c|c|c|c|c||c|c|c|c|c}
\diagbox{$\gamma$}{$h$} & 2 & 5 & 10 & 50 & 100 & 200 & [5, 2] & [5, 5] & [10, 5] & [50, 10] & [100, 50] 
\\ \hline
$10$ &
\cellcolor{green!25}{\bf 0.06} & 0.14 & 0.23 & 1.63 & 1.87 & 11.41 &
\cellcolor{green!25}{\bf 0.56} & 1.62 & 2.28 & 3.68 & 9.74
\\
$20$ &
\cellcolor{green!25}{\bf 0.14} & 0.67 & 0.21 & 2.99 & 11.85 & 63.03 &
8.86 & \cellcolor{green!25}{\bf 1.34} & 5.64 & 14.32 & 59.28
\\
$50$ &
\cellcolor{green!25}{\bf 0.11} & 2.27 & 1.96 & 7.02 & 21.65 & 110.25 &
121.30 & 21.78 & \cellcolor{green!25}{\bf 3.26} & 82.44 & 158.09 
\\
$100$ &
3.68 & \cellcolor{green!25}{\bf 1.90} & 3.03 & 11.46 & 51.63 & 119.40 &
oot & oot & \cellcolor{green!25}{\bf 222.12} & oot & oot
\\
$200$ &
48.17 & \cellcolor{green!25}{\bf 23.10} & 53.17 & 30.89 & 165.99 & 301.71 &
oot & oot & oot & oot & oot  
\\
$500$ &
oot & 70.65 & 72.09 & \cellcolor{green!25}{\bf 12.01} & 33.91 & 371.65 &
oot & oot & oot & oot & oot  
\end{tabular}
\caption{Performance results in terms of computational time [sec] varying the number of hidden neurons $h$ and the radius $\gamma$ of the domain $\mathcal{D}$. 
The fastest outcomes for one- and two hidden-layer LNN are highlighted; oot indicates timeout. 
}
\label{tab:results-compare-neurons-domain}
\end{table*}

\section{Verification of LNNs using SMT solving}
\label{sec:verifier}

Satisfiability modulo theories (SMT) comprises diverse
methods for deciding the satisfiability of first-order logic
formulae. SMT solvers combine combinatorial and symbolic
algorithms which,
unlike common numerical solvers and optimisers,
provide formal guarantees about their results
that are equivalent to those of analytical proofs.
We employ SMT solving for deciding whether
a neural network is a Lyapunov function, or for
finding counterexamples otherwise, 
which is the core of our verifier architecture. 

Deciding whether a neural network $V$ is a Lyapunov function
for a system with equilibrium (w.l.o.g.) $x_e = 0$ and within
the domain $\mathcal{D}$ amounts to deciding the formula 
\begin{equation}
  \forall x \colon (x \in \mathcal{D} \land x \neq 0) \Rightarrow (\dot{V}(x) < 0 \land V(x) > 0).\label{eq:verification}
\end{equation}
The formula verifies the Lyapunov
conditions (see \Sec \ref{sec:intro});
note that we omit the condition $V(0) = 0$ because we
guarantee it in advance by selecting biases (see \Sec \ref{sec:CEGIS-LNN}).
If the formula is true, then $V$ is a valid Lyapunov function.
However, solving large and quantified formulae can be
hard in general. For this reason, we rephrase the problem
into smaller and existential satisfiability queries
that can be SMT solved efficiently in practice.
To this end, we consider the dual falsification problem,
which is a standard approach in formal verification. 

The falsification problem is the logical
negation of the verification problem in \eqref{eq:verification}
and corresponds to the formula
\begin{multline}
    \exists x \colon
    \overbrace{(x \in \mathcal{D} \land x \neq 0 \land \dot{V}(x) \geq 0)}^{\varphi_1}
    \lor \\
    \underbrace{(x \in \mathcal{D} \land x \neq 0 \land V(x) \leq 0)}_{\varphi_2}, \label{eq:falsification}
\end{multline}
which determines whether a counterexample exists. 
If the falsification formula is true then $V$ is invalid.
Equivalently, it is true if
either $\varphi_1$ or $\varphi_2$ are satisfiable, independently
of one another.
Thanks to this,
our verifier checks each of the two sub-formulae with an
independent satisfiability
query to an SMT solver.
If the query for $\varphi_1$
produces a satisfying assignment $c_1$ or that for $\varphi_2$
produces a satisfying assignment $c_2$,
then either of $c_1$ and $c_2$ constitutes a counterexample.
The verifier adds either or both counterexamples
to the samples set $S$ and the CEGIS loop continues.
Conversely, if both queries determine that the respective formulae are unsatisfiable, 
then this means that $V$ is a valid Lyapunov function, 
and the overall loop terminates successfully.

The communication between verifier and learner is crucial
for converging quickly.
Adding one or two counterexamples at a time might make the
learner overfit each of them, which often induces long 
sequences of counterexamples that are close to one another. 
For this reason, after producing every counterexample $c$, 
we augment the samples set with a number of random additional
points from a neighbourhood of $c$.
While these additional points do not necessarily
satisfy \eqref{eq:falsification}, this expedient 
enhances the information sent to the learner, 
helps it to generalise a Lyapunov function more quickly, 
and does not hinder its overall soundness.  

The expressions for $\mathcal{D}$, $V$, and $\dot{V}$
determine the predicates that appear in
the formulae $\varphi_1$ and $\varphi_2$, 
and hence the theory for the SMT solver, which the verifier has to adopt. 
Note that, ultimately, the class of $V$ and $\dot{V}$ functions 
are determined by the activation functions $\sigma$ and by the vector field $f$. 
We experiment with polynomial $\mathcal{D}$ and $\sigma$, 
and with several polynomial systems of ODEs (see \Sec \ref{sec:case-studies}) which, 
in their turn, induce polynomial $V$ and $\dot{V}$. 
For this reason, we employ SMT solving over 
non-linear real arithmetic (NRA) which, for polynomials, is sound and complete \cite{JovanovicM12}. 
Consequently, our verifier is correct both 
when it determines that $V$ is valid and when it provides counterexamples. 
Notably, the cognate $\delta$-complete method satisfies the earlier
condition (soundness), but may in general produce spurious counterexamples \cite{ChangRG19}.  

\section{Case Studies and Experimental Results}
\label{sec:case-studies}

We provide a portfolio of benchmarks and evaluate our
  method experimentally.
In our experiments we use
quadratic activations,
i.e., $\sigma(p) = p^2$, 
but our framework supports any polynomial. 
Since CEGIS may not terminate in general, 
we set a timeout 
 of 100
  iterations and limit the verification time to 30 seconds. 
We use $\varepsilon=0.01$ for the loss function (see \Sec \ref{sec:learner}) and set $a$ to be proportional to the domain and  the system dynamics. 
Specifically, we consider the largest magnitude point $s_M$ in $S$  and compute its value $f(s_M)$;
we then approximate $a = f(s_{M})^{-1}$  to the closest power of 10. 
As for the verifier, we sample 20 additional random points for every counterexample (see \Sec \ref{sec:verifier}).
We use PyTorch to implement and train LNNs, and Z3 \cite{de2008z3} to verify them. 

We test the performance of our method varying 
the depth and width of the LNN and the input domain $\mathcal{D}$. 
We consider the system in Eq. \eqref{eq:parrilo} and six spherical domains, with radius $\gamma$ ranging from 10 to 500.
The LNN is either composed by a single hidden layer with the number of neurons 
$h_1$ in $\{2, 5, 10, 50, 100, 200 \}$,  
or by two layers with neurons  
$(h_1, h_2)$ within $\{ (5, 2), (5,5), (10, 5), (50, 10), (100, 50) \}$. 
The outcomes of the computational times are reported in Table \ref{tab:results-compare-neurons-domain}. 
Intuitively, enlarging the domain makes the search of a valid Lyapunov function harder,   
as the verifier understandably suffers from larger domains.  
Our results show that a single-layer fits best the synthesis of Lyapunov functions for the system in Eq. \eqref{eq:parrilo}: 
a quadratic activation function is sufficiently expressive, 
and surely has the least computational overhead. 
Furthermore, we highlight a dependence between the size of the LNN and the domain diameter:  
a small number of neurons might not provide the necessary flexibility to the NN to compute a Lyapunov function over a large domain. 
For this reason, utilising a multi-layer network is promising, although it must be still optimised towards generalisation in learning and towards scalability in verification.      
%



We compare our approach against {\em Neural Lyapunov Control} (NLC) \cite{ChangRG19}, which is similar to our method, 
against
a {\em constraint-based synthesis} (CBS) method \cite{tacas20},
and against SOSTOOLS \cite{papachristodoulou2018sostools}. 
We challenge our procedure by considering systems that do not admit a global polynomial Lyapunov function 
and , as in \cite{goubault2014finding}, we focus on the positive orthant of the state space. 
Data points close to $x_e$ represent a numerical and analytical challenge to the NLC algorithm. 
Thus, as per \cite{ChangRG19}, we remove a sphere around the origin from the domain, hence considering 
$\mathcal{D}(\rho, \gamma)$ = $\{ x_i \geq 0$, $\forall i $, $\rho \leq ||x||_2 \leq \gamma \}$, where $\rho$ and $\gamma$ represent the radii of inner and outer spheres, respectively.
A Lyapunov function valid on such a domain proves practical (or Lagrange) stability, 
which is weaker than Lyapunov asymptotic stability obtained in our work.    
We report results in terms of computational time and maximum $\gamma$ in Table \ref{tab:nl-results}.  
We consider the system in Eq. \eqref{eq:parrilo}, together with the following models \cite{goubault2014finding}: 
\begin{equation}
\begin{cases}
\dot{x} = -x + 2x^2y \\
\dot{y} = - y, 
\end{cases}
\label{cs:non-poly-square}
\end{equation}
\begin{equation}
\begin{cases}
\dot{x} = -x \\
\dot{y} = - 2y + 0.1 x y^2 + z \\
\dot{z} = -z -1.5 y, 
\end{cases}
\label{cs:non-poly-3dim-easy}
\end{equation}
\begin{equation}
\begin{cases}
\dot{x} = -3x -0.1 x y^3 \\
\dot{y} = - y + z \\
\dot{z} = -z.  
\end{cases} 
\label{cs:non-poly-3dim-hard}
\end{equation}


\begin{table*}[ht]
\centering
\begin{tabular}{c||c|c|c||c|c|c||c|c|c||c|c}
Test & LNN Total & LNN Ver. & LNN & NLC Total & NLC Ver. & NLC & CBS & CBS Ver. & CBS & SOS & SOS
\\ 
Eq. \# & Time [sec] & Time [sec] & $\gamma$ & Time [sec] & Time [sec] & Domain & Time [sec] & Time [sec] & $\gamma$ & Time [sec] & $\gamma$ 
\\ 
\hline
\eqref{eq:parrilo} & 12.01 & 1.28 & 500 & 6.28 & 0.29 & $\mathcal{D}(0.1, 1)$ & 0.22 & 0.08 & 1 & 6.67 & 800
\\
\eqref{cs:non-poly-square} & 0.29 & 0.08 & 100 & 5.45 & 0.22 & $\mathcal{D}(0.1, 1)$ & 0.30 & 0.09 & 1 & 7.76 & 25
\\
\eqref{cs:non-poly-3dim-easy} & 0.32 & 0.29 & 1000 & 54.12 & 23.70 & $\mathcal{D}(0.1, 1)$ & 2.22 & 0.58 & 1 & 11.80 & oot
\\
\eqref{cs:non-poly-3dim-hard} & 33.27 & 33.11 & 1000 & 37.80 & 13.45 & $\mathcal{D}(0.1, 1)$ & 0.42 & 0.09 & 1 & 9.65 & oot
\end{tabular}
\caption{
Comparison between proposed approach (LNN), CBS and NLC approaches, and SOSTOOLS: 
total computation time, verification time, and domain
width. Timeouts are indicated with oot. 
}
\label{tab:nl-results}
\end{table*}

Both NLC and CBS successfully synthesise Lyapunov functions for domains of radius $\gamma = 1$ but  time out with larger $\gamma$.
Our method shows faster results and synthesises over wider domains: 
 we successfully synthesise Lyapunov function with domains of radius $\gamma \geq 100$ for all models. 
In three out of four benchmarks we are faster than NLC, whilst coping with wider domains. 
SOSTOOLS synthesises Lyapunov functions numerically but does not provide
a sound verification check; for this reason, we pass its result
to Z3 for computing its validity domain. 
Whilst SOSTOOLS is fast, it generally returns Lyapunov functions with ill-conditioned coefficients
that affect the verification step, which times out in two of the case studies.

Neural networks can be regarded as templates: 
every hidden neuron represents a single quadratic instance, whereas more layers generalise the LNN to higher-order polynomials. 
However, we have demonstrated that 
LNNs have superior performance with respect to classic  template-based methods (\ie, CBS).
Besides, the choice of polynomial $\sigma$ maintains the intuition of Lyapunov functions as energy-like functions for ODEs, 
whilst remaining within the range of functions that are verifiable algorithmically. 

\section{Conclusion}
\label{sec:conclusions}

We have proposed a neural approach to automatically
synthesise provably correct Lyapunov functions for polynomial
systems.
We have employed a CEGIS architecture, where a learner trains
Lyapunov Neural Networks using machine learning and the verifier
validates them or finds counterexamples using SMT solving.
We have 
compared our method against alternative approaches on 4 case studies.
Our method has computed Lyapunov functions faster than NLC, over wider
domains than NLC and CBS, and giving stronger guarantees
than NLC and SOSTOOLS. 
Our method offers ease of implementation, because learner and
verifier use black-box machine learning and verification
techniques and are independent of one another.
However,
CEGIS can in general suffer from 
unreasonably (or infinitely) many iterations.

We have tackled the stability analysis of autonomous systems. 
Automated control synthesis requires considering additional inputs variables, 
and the performance of the verifier is sensitive to the system dimensionality: 
as such, scalable verification of neural networks is subject of active research and matter for future work.


\bibliographystyle{IEEEtranS} 
\bibliography{LyapNN}

\end{document}